# Quantized Non-Volatile Nanomagnetic Synapse based Autoencoder for Efficient Unsupervised Network Anomaly Detection

Muhammad Sabbir Alam[1], Walid Al Misba[1], Jayasimha Atulasimha[1,2*]

[1] Dept. of Mechanical and Nuclear Engineering, Virginia Commonwealth University, Richmond, VA, 23284, USA
[2] Dept. of Electrical and Computer Engineering, Virginia Commonwealth University, Richmond, VA, 23284, USA

[*]Email: jatulasimha@vcu.edu

**Abstract:** In the autoencoder based anomaly detection paradigm, implementing the autoencoder in edge devices capable of learning in real-time is exceedingly challenging due to limited hardware, energy, and computational resources. We show that these limitations can be addressed by designing an autoencoder with low-resolution non-volatile memory-based synapses and employing an effective quantized neural network learning algorithm. We propose a ferromagnetic racetrack with engineered notches hosting a magnetic domain wall (DW) as the autoencoder synapses, where limited state (5-state) synaptic weights are manipulated by spin orbit torque (SOT) current pulses. The performance of anomaly detection of the proposed autoencoder model is evaluated on the NSL-KDD dataset. Limited resolution and DW device stochasticity aware training of the autoencoder is performed, which yields comparable anomaly detection performance to the autoencoder having floating-point precision weights. While the limited number of quantized states and the inherent stochastic nature of DW synaptic weights in nanoscale devices are known to negatively impact the performance, our hardware-aware training algorithm is shown to leverage these imperfect device characteristics to generate an improvement in anomaly detection accuracy (90.98%) compared to accuracy obtained with floating-point trained weights. Furthermore, our DW-based approach demonstrates a remarkable reduction of at least three orders of magnitude in weight updates during training compared to the floating-point approach, implying substantial energy savings for our method. This work could stimulate the development of extremely energy efficient non-volatile multi-state synapse-based processors that can perform real-time training and inference on the edge with unsupervised data.

## 1. Introduction

In today's interconnected world, the security and integrity of computer networks are of paramount importance. By 2030, it is estimated that 500 billion devices will be connected to the internet [1], with a significant portion comprising Internet of Things (IoT) devices. While the rapid growth of network-based devices, applications and services offer immense convenience, it also has led to an increasing number of cyber threats and attacks [2]. The proliferation of cybercrimes and network intrusions underscores the need for developing robust solutions that can safeguard network security. Anomaly detection plays a vital role in safeguarding these networks by identifying and mitigating abnormal or malicious activities that deviate from expected patterns [3]. Detecting such anomalies in real-time is crucial to prevent potential damage, data breaches, and service disruptions [4]. As we look towards the future, where the Internet of Things (IoT) and edge computing gain prominence, the need for efficient and effective anomaly detection becomes even more critical.

Autoencoders have emerged as a promising approach for anomaly detection in unlabeled network traffic samples [5]-[7]. These neural network architectures are capable of learning meaningful representations of data by training on unlabeled examples. The essence of an autoencoder lies in its ability to encode input data into a lower-dimensional latent space and then decode it back to reconstruct the original input. By leveraging the power of deep learning, autoencoders excel at capturing and representing complex patterns and structures in data, making them well-suited for unsupervised learning inspired anomaly detection tasks. However, implementing autoencoders for real-time anomaly detection on resource-constrained edge devices poses significant challenges. Edge devices, such as internet of things (IoT) devices and embedded systems, typically have limited hardware capabilities, computational resources, and energy constraints [8], [9]. These constraints hinder the deployment of sophisticated deep learning models like autoencoders on the edge, necessitating the development of novel approaches that can meet these challenges.

The autoencoder's synaptic weight precision can be reduced using quantization techniques [10], [11], which offers an effective solution for energy and computational resource-constraint environments, such as edge devices. By employing low-resolution weight representations instead of floating-point weights, quantization reduces memory footprint, energy consumption, and latency, thus enabling efficient implementation of neural networks [10], [11]. However, traditional post training quantization methods [10], which involve applying quantized weights directly to pre-trained models, may lead to a degradation in model accuracy due to information loss caused by weight rounding. Quantization-aware training aims to address this challenge by integrating the quantization process during the training phase itself. This approach ensures that the neural network learns to adapt to the reduced precision and achieve better performance when deployed in low-precision environments [12], [13].

Besides quantization, further reduction in computational resources can be achieved using in-memory computing [14], [15] platform, where the data can be stored and processed directly within the main memory, eliminating the need for frequent data transfers between memory and processor, as commonly observed in the von-Neumann computing paradigm [16]. Therefore, this approach leads to faster and more efficient data processing. Spintronic memory devices offer a promising solution for implementing in-memory neural networks because of their non-volatility, higher energy efficiency, faster speed, dense footprint, and CMOS compatibility [16]-[20]. For instance, spintronic magnetic domain wall (DW) devices [17], [18] possess all these desirable properties. However, due to small on/off ratio of magnetic tunnel junctions (MTJs), which can be at most 7:1 at room temperature [21], DW devices that encode information in the resistance of MTJs have drawbacks including inherent low-resolution as well as stochastic nature [22], [23]. These imperfect device characteristics typically have a negative impact on neural network accuracy. Nevertheless, many research studies have demonstrated that modified training algorithms can retain high accuracy in analog low-resolution device-based neural networks [24]-[26].

This study introduces a novel approach to efficient anomaly detection in edge nodes through the utilization of a low-resolution quantized DW synapse-based autoencoder. The autoencoder's synapses are designed using a ferromagnetic racetrack device hosting a magnetic DW and having engineered notches at a regular interval. The synaptic weights, limited to five states, are controlled by applying spin orbit torque (SOT) current pulses in the heavy metal layer underneath the magnetic racetrack. Extensive micromagnetic simulations are conducted in the presence of room temperature thermal noise to assess the stochastic variations in each memory state of the DW device. To enhance the autoencoder's performance, a hardware-aware training algorithm is employed, inspired by neural network training with low-precision weights. In this algorithm, weight gradients are accumulated in a separate high-precision memory before being quantized and programmed into the low-resolution devices in the analog domain. Our proposed method is tested on the NSL-KDD [27] dataset, which is an exemplary dataset that has been widely used for evaluating the performance of intrusion detection methods. This dataset provides a realistic and challenging environment for assessing the efficacy of anomaly detection algorithms, allowing researchers to determine the robustness and generalizability of their proposed methods.

The contributions of our proposed method can be summarized as follows: Firstly, we explore the feasibility of utilizing unsupervised learning-based deep neural network training with quantization-aware training. To the best of our knowledge, quantization-aware training has not been extensively explored in the context of unsupervised learning. The results of our study demonstrate that training with 5-state synaptic weights yields superior performance compared to training with 2-state or 3-state synaptic weights. Secondly, we demonstrate that, by employing the effective hardware-aware training algorithm, our proposed autoencoder model achieves a high level of testing accuracy in anomaly detection, surpassing the testing accuracy obtained when using floating-point trained weights. The combination of quantization noise and device stochasticity acts as an effective regularization process, thereby improving the testing accuracy of the proposed autoencoder model. Furthermore, our DW-based approach showcases a significant reduction of at least three orders of magnitude in weight updates during training in contrast to the floating-point approach, indicating substantial energy conservation benefits inherent to our method. While our study focuses on the use of the DW device as an illustrative example, the insights gained from our research can be extended to other non-volatile multistate memory technologies. This exploration opens up avenues for implementing quantized autoencoders in anomaly detection, leading to improved efficiency and effectiveness.

The subsequent sections of the paper are structured as follows: Section 2 explores the related research. Section 3 delves into the data and the data preprocessing steps. Section 4 provides comprehensive information regarding the proposed quantized autoencoder-based anomaly detection, including details on the autoencoder model, quantization-aware training process, anomaly detection workflow, and algorithm. Section 5 presents the design of the DW synapse and the micromagnetic simulations required for the DW synapse-based autoencoder. Section 6 investigates the performance of anomaly detection using the proposed method. Finally, Section 7 concludes the paper by discussing the outcomes of this study and outlining potential future directions.

## 2. Related Work

In recent years, the increasing adoption of machine learning approaches for anomaly detection has been driven by the limitations and high costs associated with conventional signature-based intrusion detection techniques [28]. These traditional methods prove inadequate in effectively detecting zero-day attacks, which are characterized by their unknown and previously unseen nature [29]. Various supervised learning-based classification algorithms and hybrid models combining multiple algorithms have been explored to identify network anomalies and detect attacks with high accuracy. Notable algorithms include Support Vector Machine (SVM), Decision Tree (DT), Naive Bayes Network (NB), Naive Bayes Tree (NBTree), J48, Fuzzy logic, and Artificial Neural Networks (ANN) [27], [30]-[32]. However, the effectiveness of these algorithms hinges on accurate labels and balanced training data [33]. The availability of such data, particularly in the realm of network intrusion detection, is limited due to factors like privacy concerns and data confidentiality [34]. To overcome this constraint, researchers have turned to unsupervised learning techniques, such as anomaly detection algorithms based on autoencoders [6], [7]. More recently, studies have been conducted on unsupervised deep learning circuits using memristors to enable real-time anomaly detection with autoencoders on low-power devices [35].

In the emerging field of spintronic memory devices, researchers have made significant advancements in leveraging their properties for efficient computing. It has been shown that non-volatile nanomagnetic devices can be controlled efficiently using voltage-controlled magnetic anisotropy (VCMA) [36]-[38], voltage-induced strain [39]-[43], current control [44], [45], and a combination of both current and voltage control [23], [46], [47]. Studies have demonstrated that, despite having imperfect device characteristics, nanomagnetic devices can be implemented as multistate synapses for deep neural networks [23], [24]. However, for quantized neural network implementation, weight gradients need to be stored in high-precision memory during the training phase to retain high accuracy [11], [48]. In recent studies, it has been observed that by incorporating an effective quantization-aware training algorithm, stochastic and extremely low-resolution (less than 3-bit) DW devices can achieve comparable accuracy levels to floating-point precision neural networks [24].

While the feasibility and implementation of various deep neural network architectures using emerging spintronic memory devices are actively being studied, the implementation of autoencoders using nanomagnetic devices remains relatively unexplored. This research gap serves as the motivation to explore the connection between the fields of unsupervised autoencoder-based anomaly detection and spintronic nanomagnetic synapse technology.

## 3. Data and Preprocessing

In this section, we explore the characteristics of the NSL-KDD dataset and outline the preprocessing steps undertaken to prepare the data for further analysis.

### 3.1 NSL-KDD Data

The NSL-KDD dataset is derived from the KDD Cup 1999 dataset, which represents a comprehensive collection of network traffic data containing both normal and various attack instances [27]. Within the NSL-KDD dataset, two distinct sets of data exist: KDDTrain+ and KDDTest+. The training data (KDDTrain+) consists of 125,973 packets, each categorized into one of the 23 distinct data types. These types include malicious categories such as Ipsweep, Guess_passwd, Warezclient, Neptune, Multihop, Perl, Smurf, Phf, Rootkit, Imap, Loadmodule, Portsweep, Nmap, Back, Pod, Spy, Land, Warezmaster, Satan, Buffer_overflow, Teardrop, Ftp_write, as well as the category labeled Normal [49]. Among these, 58,630 packets in the KDDTrain+ dataset are labeled as malicious, while the remaining 67,343 packets represent the normal packets. As for the KDDTest+ dataset, it comprises a total of 22,544 packets, with 12,833 packets categorized as malicious and the remaining 9,711 packets labeled as normal.

TABLE 1
Features and Data Types of NSL-KDD Dataset

| No | Attribute | Type | No | Attribute | Type |
|---|---|---|---|---|---|
| 1 | duration | int64 | 22 | is_guest_login | int64 |
| 2 | protocol_type | object | 23 | count | int64 |
| 3 | service | object | 24 | srv_count | int64 |
| 4 | flag | object | 25 | serror_rate | float64 |
| 5 | src_bytes | int64 | 26 | srv_serror_rate | float64 |
| 6 | dst_bytes | int64 | 27 | rerror_rate | float64 |
| 7 | land | int64 | 28 | srv_rerror_rate | float64 |
| 8 | wrong_fragment | int64 | 29 | same_srv_rate | float64 |
| 9 | urgent | int64 | 30 | diff_srv_rate | float64 |
| 10 | hot | int64 | 31 | srv_diff_host_rate | float64 |
| 11 | num_failed_logins | int64 | 32 | dst_host_count | int64 |
| 12 | logged_in | int64 | 33 | dst_host_srv_count | Int64 |
| 13 | num_compromised | int64 | 34 | dst_host_same_srv_rate | float64 |
| 14 | root_shell | int64 | 35 | dst_host_diff_srv_rate | float64 |
| 15 | su_attempted | int64 | 36 | dst_host_same_src_port_rate | float64 |
| 16 | num_root | int64 | 37 | dst_host_srv_diff_host_rate | float64 |
| 17 | num_file_creations | int64 | 38 | dst_host_serror_rate | float64 |
| 18 | num_shells | int64 | 39 | dst_host_srv_serror_rate | float64 |
| 19 | num_access_files | int64 | 40 | dst_host_rerror_rate | float64 |
| 20 | num_outbound_cmds | int64 | 41 | dst_host_srv_rerror_rate | float64 |
| 21 | is_host_login | int64 | | | |

Each packet in the NSL-KDD dataset consists of 41 features that are identical to those in the KDD Cup 1999 dataset. Table 1 provides an overview of the 41 features along with their corresponding data types. The 42$^{nd}$ feature represents the data label (normal/attack) [49]. However, for training purposes, the packets do not include the data label. Among the selected features, 38 are numeric (int64/float64) and the remaining 3 are categorical (object) variables.

### 3.2 Preprocessing Steps

Prior to the training phase, preprocessing is applied to all packets in the dataset. First, the categorical features are converted into numerical representations. Specifically, the second position (protocol/type), third position (service), and fourth position (flag) contain categorical data. The three categorical features are modified to one-hot encoded numerical values. For instance, in the "protocol type" feature, the strings "tcp", "udp", and "icmp" are replaced with their respective one-hot encoded representations: [1 0 0], [0 1 0], and [0 0 1]. As a result, the three categorical features in the NSL-KDD dataset, namely "protocol type," "service," and "flag," which have 3, 70, and 11 distinct strings respectively, are transformed into a total of 84 features. This one-hot encoding process leads to a combined total of 122 features, including the selected original 38 numeric features.

Additionally, the dataset undergoes normalization to ensure consistency. Each feature vector is normalized by scaling them to fit in the range of [0; 1] based on the maximum value within that feature vector. All types of malicious data are labeled as "1," while normal data packets are labeled as "0".

## 4. Quantized Autoencoder based Anomaly Detection

### 4.1 Autoencoder

An autoencoder is a neural network architecture that employs unsupervised learning to reconstruct input data. Comprising multiple layers, including one or more hidden layers, the autoencoder maintains the same size for both the input and output layers. At the center of the network lies the bottleneck layer, which represents a compressed latent space representation of the input data. The encoder maps the input to the bottleneck layer representation, while the decoder reconstructs it in the output layer [7]. Fig. 1 illustrates the architecture of a standard autoencoder.

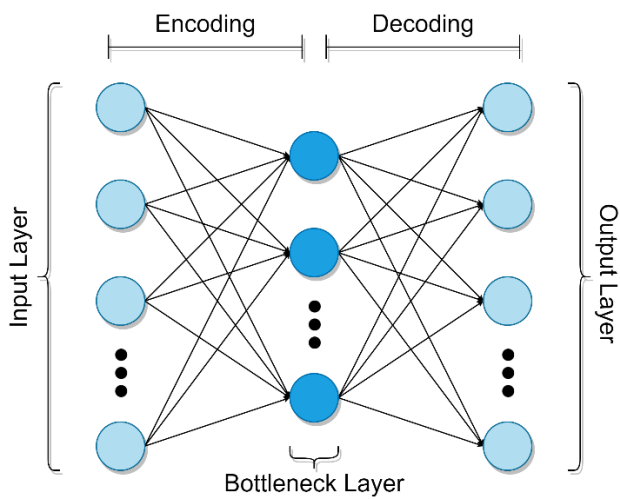

Figure 1: A standard autoencoder architecture.

In the encoding phase, an n dimensional input vector X $[x_1, x_2, x_3, \ldots, x_n]$ is mapped to hidden layer representation H. The encoding operation is expressed as equation (1):

$$H = F_1(W_1 X + b_1) \tag{1}$$

where $W_1$ is the weight matrix, $b_1$ is the bias vector, and $F_1$ denotes the encoder activation function.

In the decoding phase, the latent space representation H is mapped to reconstruct the input (X) as output R $[r_1, r_2, r_3, \ldots, r_n]$. The decoding operation is expressed as equation (2):

$$R = F_2(W_2 Y_1 + b_2) \tag{2}$$

where $W_2$ is the weight matrix, $b_2$ is the bias vector, and $F_2$ denotes the decoder activation function.

As the autoencoder goes through training with backpropagation, the weights and biases are updated to minimize the loss function (L). The loss function is used to minimize the reconstruction error. It can be expressed as equation (3):

$$L(\Theta) = \frac{1}{n}\sum_{i=1}^{n} \|X_i - R_i(\Theta)\|^2 \tag{3}$$

where $\Theta$ is the autoencoder parameters (weights and biases). Here, L represents a mean squared error (MSE) loss function.

The reconstruction error is used to determine whether a network traffic sample is normal or malicious. During the testing phase, if a network sample shows a high reconstruction error, it is likely to be considered as a malicious packet. This is because an autoencoder trained on normal network traffic packets generally has low reconstruction error for normal data.

**4.2 Autoencoder Model**

In this study, we use an autoencoder architecture comprising five layers. The number of nodes in each layer, ranging from the input to the output layer, is [122-32-10-32-122], influenced by the model architectures investigated in [7]. However, it is important to note that our primary focus is not the impact of different autoencoder model architectures; instead, we aim to compare the performance of an autoencoder with quantized synapses to a similar autoencoder with floating-point synapses. The input features undergo a sequence of weighted sum operations and activation functions within the autoencoder model until they are reconstructed in the output layer. Specifically, we employ the sigmoid function as the activation function. The autoencoder is trained using the backpropagation algorithm in conjunction with stochastic gradient descent. This enables the adjustment of the weights and biases of the network based on the difference between the predicted output and the actual input. The performance of the autoencoder is measured using MSE as the loss function. Fig. 2 illustrates the autoencoder architecture.

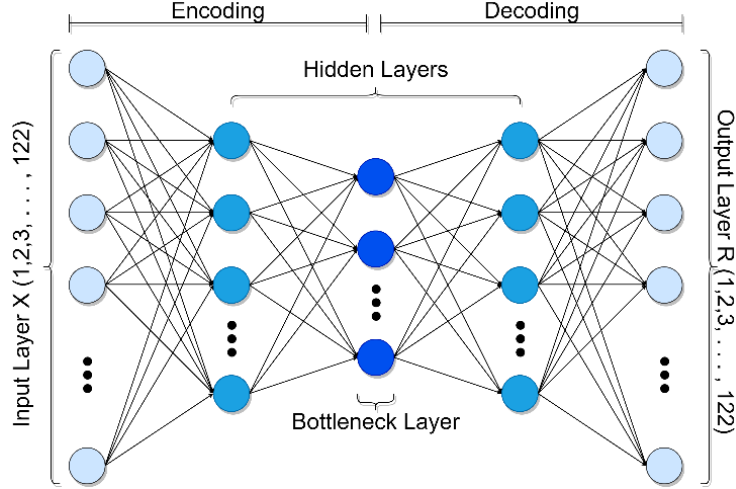

Figure 2: Proposed autoencoder model with five layers.

### 4.3 Autoencoder Quantization

Since the proposed nanomagnetic synapses have a limited number of states, the autoencoder parameters are quantized in the training and the inference stages. The quantization process can be mathematically expressed with following functions [50]:

$$\text{clip}(N_{fp}; l, h) = \max(l, \min(h, N_{fp}))$$

$$N_q = \text{round}\left(\frac{\text{clip}(N_{fp}; l, h) - l}{h - l}(n - 1)\right) \times \frac{h - l}{n - 1} + l \quad (4)$$

where $N_q$ is the quantized value, $N_{fp}$ is the full precision value, [l; h] is the quantization range, and n is the number of quantization levels.

### 4.4 Autoencoder Training Process

While KDDTrain+ and KDDTest+ carry both normal and malicious traffic samples, only the normal traffic samples from the KDDTrain+ are used for training the quantized autoencoder. Fig. 3 illustrates the workflow for quantized autoencoder based anomaly detection. In the training phase, the processed normal traffic samples are sent to the autoencoder, where the original features are encoded to a latent space representation. Next, output features are reconstructed from this latent space. The reconstruction error is assessed from the difference between the reconstructed traffic sample and the original sample. The reconstruction error is measured for all the traffic samples and the standard deviation is estimated, which acts as the threshold for detecting anomalies. In the inference phase, network traffic samples (carrying both normal and malicious samples) are sent to the trained autoencoder, and reconstruction error is calculated for each sample. The difference between the reconstruction error of a sample and the mean error of samples is referred to as the anomaly score (AS). The anomaly score is then compared to the threshold. If the anomaly score is higher than or equal to the threshold, then the traffic sample is inferred as malicious. Otherwise, the sample is inferred as benign.

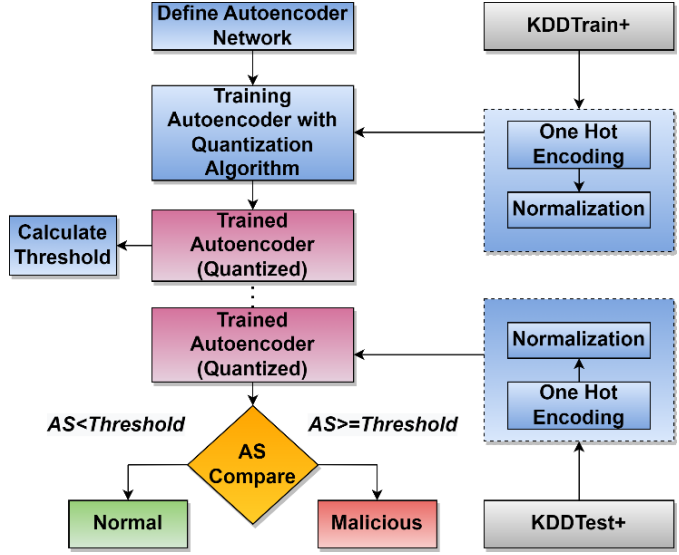

Figure 3: Schematic diagram of quantized autoencoder based anomaly detection.

Quantization-aware training of the autoencoder is performed by quantizing the weights according to (4) in the feed-forward phase and using the backpropagation algorithm based on low-precision neural network training [11]. In this process, while all weights are quantized in the forward pass, weight gradients are stored in separate high precision memory units during the backward pass to retain accuracy. However, the weight gradients need to propagate through non-differentiable quantization blocks in the backward pass. To tackle this issue, the straight through estimator (STE) is used, which provides a workaround by treating the quantization operation as a simple identity function during backpropagation [11]. This allows the gradients to be backpropagated as if the quantization were not applied.

**Algorithm:** Quantization-aware training for DW synapse-based autoencoder for network anomaly detection:

**Input:** Training dataset $\mathbf{X_0} = [X_1, X_2, X_3, \ldots, X_N]$
Testing dataset $\mathbf{T_0} = [T_1, T_2, T_3, \ldots, T_M]$

// X and T are both n dimensional vectors
// Encoder: **E**, Decoder: **D**, *Cost function: C*
// Output: Anomaly = "1", Normal = "0"
// Number of Layers: L, Learning rate: η, Noise Tolerance Margin: α

**begin**
   //Training Phase
   Initialize Weights: **W** ← Gaussian Dist. [-1, 1]
  **for** number of training iterations **do**
  //Step 1: Feed-Forward
  **for** *k = 1 to L* **do**
  $\mathbf{X_k} = \mathbf{X_{k-1} W_k}$
  **end**
  $\mathbf{Y} = \mathbf{X_L} = \mathbf{D(E(X))}$
  //Step 2: Compute Gradients
  Compute gradient $\mathbf{G_L} = \frac{\partial C}{\partial X_L}$ from $\mathbf{X_L}$ and $\mathbf{X_0}$

**for** *k = L to 1* **do**
$G_{k-1} = G_k W_k$
$\Delta W_k = \eta G_k X_{k-1}$
**end**
//Step 3: Weight Update
**for** *k = 1 to L* **do**
$W_{k,fp}(t+1) \leftarrow$ Update $(W_{k,fp}(t), \Delta W_k)$
$W_{k,quant}(t+1) \leftarrow$ Quantize (Clip $(W_{k,fp}(t+1), -1, 1)$, QL)
//QL = Number of states for Quantization
  **if** $|W_k(t) - W_{k,quant}(t+1)| > \alpha$ **then**
    $W_k(t+1) \leftarrow$ Program $(W_k(t), W_{k,quant}(t+1))$
  **end**
**end**
**end**
//Threshold Calculation from Training dataset
**for** *j = 1 to N* **do**

$$D_j = \sqrt{\sum_{i=1}^{n} \|X_{ji} - Y_{ji}\|^2}$$

**end**

$$D_m = \frac{1}{N} \sum_{j=1}^{N} D_j$$

Threshold, $D_{SD} = \sqrt{\frac{\sum_{j=1}^{N} \|D_j - D_m\|^2}{N}}$

// Inference Phase
$Z = T_L = D_{quant}(E_{quant}(T))$
**for** *j = 1 to M* **do**

$$F_j = \sqrt{\sum_{i=1}^{n} \|T_{ji} - Z_{ji}\|^2}$$

**end**
**for** j = 1 to M **do**
$\Delta = |F_j - D_m|$
  **if** $\Delta >= D_{SD}$ **then**
    $T_j$ is an anomaly
    insert $T_j$ to Anomaly Set
  **else**
    $T_j$ is not an anomaly
    insert $T_j$ to Normal Set
  **end**
**end**
**end**

## 5. Implementing Quantized Synapses with Non-Volatile Memory

### 5.1 DW Synapse Design

In this section, we discuss the proposed low-resolution (quantized) DW synapse design for the autoencoder neurons. Being non-volatile, the DW memory retains data for a long time even in absence of power. We design the device by simulating a thin ferromagnetic racetrack with five engineered notches where DW positions can be controlled with current pulse. The racetrack has a dimension of 560 nm × 60 nm × 1 nm. Along the racetrack, engineered notches are incorporated at a regular interval of 100 nm. However, the first and the fifth notches are positioned at 80 nm and 480 nm respectively. Fig. 4 illustrates the design of the nanomagnetic DW-based synaptic device.

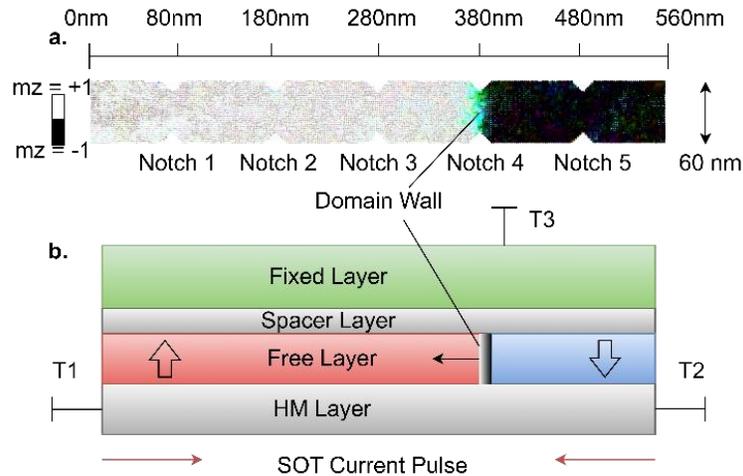

Figure 4: Schematic diagram of magnetic domain wall non-volatile synapses driven by SOT current pulses: (a) A sample of micromagnetic simulations showing pinned position of the domain wall (b) Configuration of DW device with 5 notches.

To control the DW position, current pulses with specific amplitude and fixed pulse duration are applied across the heavy metal layer. The current pulses generate SOT, which acts on the magnetic racetrack above it. By varying the number and the direction of current pulses, DW can be moved to different positions.

An MTJ is formed by combining the racetrack (free layer), an insulator (MgO tunneling layer), and a ferromagnetic material (reference layer) as seen in Fig. 4b. The MTJ is used to read the state of the racetrack's magnetization. DW positions are encoded as the conductance of MTJ, thereby creating a synapse that can be programmed with current pulses.

### 5.2 Micromagnetic Simulations

Extensive micromagnetic simulations are performed using mumax3 [51], which simulate the magnetization dynamics of the DW in the magnetic racetrack while considering thermal noise at room temperature (300 K). The simulations provide insights into the evolution of the DW synapse. Table 2 presents the parameters employed for the micromagnetic simulation.

TABLE 2
Parameters for the Micromagnetic Simulations [23]

| Symbol | Description | Value |
| --- | --- | --- |
| $\alpha$ | Damping Parameter | 0.015 |
| D | DMI Constant | 0.0006 Jm$^{-2}$ |
| $A_{ex}$ | Exchange Constant | $2 \times 10^{-11}$ Jm$^{-1}$ |
| Ψ | Gilbert Damping | 0.03 |
| $K_u$ | Perpendicular Magnetic Anisotropy (PMA) | $7.5 \times 10^5$ Jm$^{-3}$ |
| $M_s$ | Saturation Magnetization | $10^6$ Am$^{-1}$ |
| $\lambda_s$ | Saturation Magnetostriction | 250 ppm |
| T | Temperature | 300 K |

Fig. 4a illustrates the micromagnetic configuration of the racetrack's free layer. A fixed amplitude current pulse of $85 \times 10^9$ A/m$^2$ with a fixed duration of 0.5 ns is applied through the heavy metal layer to initiate the DW depinning from its initial pinned position and move it towards the intended adjacent notch. However, due to the DW tilting caused by the presence of Dzyaloshinskii-Moriya interaction (DMI) [52] and thermal noise, the DW exhibits significant stochastic motion when driven by the SOT current pulses. As a result, the DW might get pinned at a different notch position rather than getting pinned at the intended specific notch position after the SOT current pulses are applied. Fig. 5 illustrates the probabilistic distribution of the DW positions due to stochastic variation in the DW motion. The equilibrium pinned positions of the DWs are used to calculate the conductance of the MTJ using the subsequent equation (5) [24]:

$$G_{Synapse} = \frac{G_{max} + G_{min}}{2} + \frac{G_{max} - G_{min}}{2} <m_z> \quad (5)$$

where $<m_z>$ denotes the average magnetization moment of the ferromagnetic racetrack along the z-direction. The magnetization of the reference ferromagnetic layer is assumed to point upward in the +z-direction. $G_{max}$ and $G_{min}$ respectively denote the maximum and minimum conductance of the synaptic device.

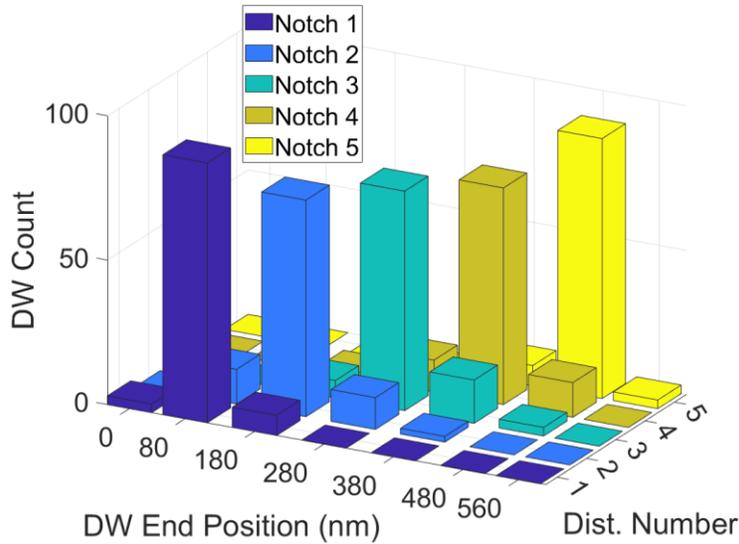

Figure 5: Probabilistic distribution of DW positions and DW counts after SOT current pulses are driven to move the DW to target notches from adjacent notch positions.

The conductance of the DW device represents the weights, which corresponds to the position of the DW. However, this means that only positive weights can be attained using the DW device. To address the need for weight updates in both positive and negative directions, a circuit model illustrated in Fig. 6 is designed [24]. This model utilizes two separate rows connected to a single column (bit line or BL) in the crossbar. These rows are supplied with opposite polarity voltages and are responsible for connecting the synaptic devices and parallel conductance ($G_p$). To enable read and write operations, additional components such as the write and read word lines (WWL/RWL) and source line (SL) are introduced (as shown in Fig. 6). It is worth noting that for simplicity, the WWL for the parallel conductance is not shown in the figure. To perform column sum read/write operations, the RWL or WWL is activated accordingly. When programming a device, the WWL is activated, and the SL and BL are adjusted to high or low levels based on the direction and number of current pulses.

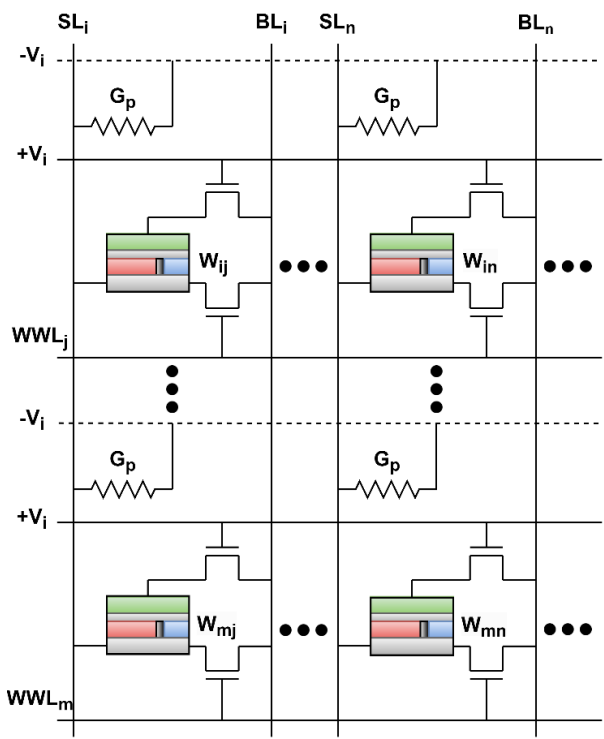

Figure 6: Implementation of autoencoder in crossbar architecture with nonvolatile DW memory-based synapses.

## 6. Results

### 6.1 Performance Measures

We employ accuracy, precision, true positive rate (TPR), and F1 score to assess the performance of the autoencoder model. These performance measures can be expressed using four quantities: True Positive (TP), True Negative (TN), False Positive (FP), and False Negative (FN). TP represents the number of correctly classified malicious samples, TN represents the number of correctly classified normal samples, FP represents the number of normal samples incorrectly classified as malicious samples, and FN represents the number of malicious samples incorrectly classified as normal samples.

Accuracy quantifies the overall correctness of the classification model and represents the ratio of correctly classified packets to the total number of packets. It is represented by the following equation (6):

$$\text{Accuracy} = \frac{TP + TN}{TP + TN + FP + FN} \tag{6}$$

Precision quantifies the proportion of correctly classified malicious samples out of all samples predicted as malicious. It is expressed as equation (7):

$$\text{Precision} = \frac{TP}{TP + FP} \tag{7}$$

TPR measures the proportion of correctly classified malicious samples out of all malicious packets. It is expressed as equation (8):

$$\text{TPR} = \frac{TP}{TP + FN} \tag{8}$$

The F1 score is a harmonic mean of precision and TPR, providing a balanced measure of both metrics. It is expressed as equation (9):

$$\text{F1 score} = \frac{2 \times \text{Precision} \times \text{TPR}}{\text{Precision} + \text{TPR}} \tag{9}$$

**6.2 Results for Quantized Autoencoder**

In this section, we compare the performance of the proposed quantized autoencoder with 2, 3, and 5-state quantization levels to a structurally identical autoencoder with full precision floating-point (32-bit) weights. The performance evaluation is based on testing accuracy, precision, TPR, and F1 score. Fig. 7 illustrates these performance metrics for the autoencoder with 2, 3, 5-state, and full precision floating-point synapses.

Fig. 7a demonstrates that employing only 2 quantization levels in the autoencoder leads to random fluctuations in the resulting accuracy as the training progresses through successive epochs. Thus, training with 2 quantization levels shows occasional high accuracy with extended training cycles; however, predicting the number of epochs required to achieve high accuracy remains challenging since the accuracy does not converge over time. Similarly, training the autoencoder with 3 quantization levels yields a comparable outcome, showing random fluctuations in accuracy with extended training cycles. However, the fluctuation pattern is less random compared to the previous case. On the other hand, training the autoencoder with 5 quantization levels demonstrates a more deterministic accuracy progression, with accuracy gradually increasing over the epochs. Notably, across all epochs, the accuracy for training with 5-state weights is comparable to the accuracy for training with full precision weights. Similar conclusions can be drawn for the remaining performance metrics illustrated in Fig. 7b, Fig. 7c, and Fig. 7d.

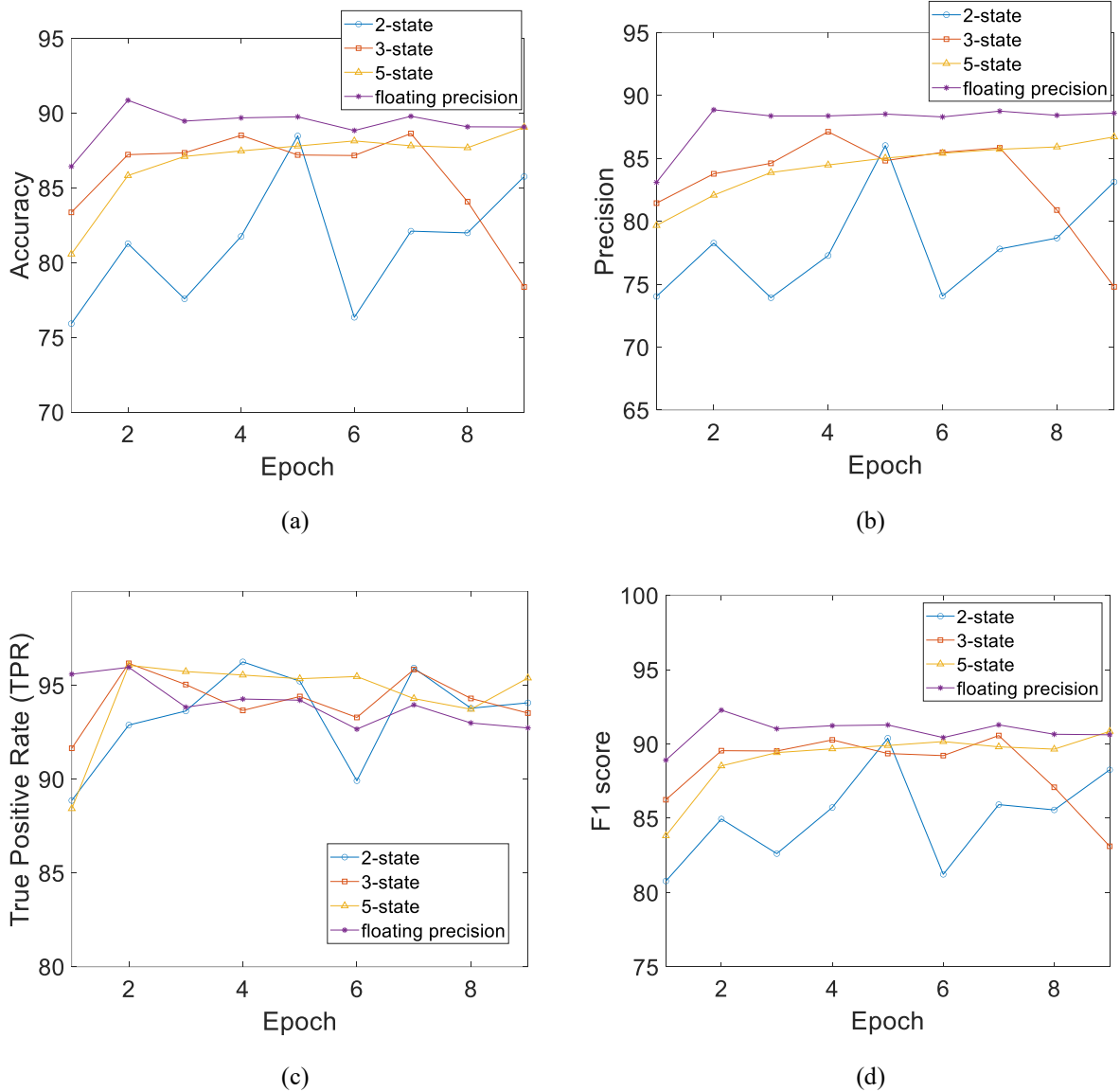

Figure 7: Anomaly detection testing (a) accuracy (b) precision (c) TPR and (d) F1 score for autoencoder with different state (2, 3, 5-state and floating precision) synaptic weights.

From the results demonstrated in Fig. 7, it can be inferred that training with 5 quantization levels shows comparable performance metrics to training with full precision weights, even though 5-state quantization requires significantly lower number of computations. While the limited number of quantized states is usually regarded as detrimental to training and testing accuracy, quantization-aware training can leverage this quantization noise to generate a significant improvement in accuracy compared to the accuracy obtained by floating-point trained weights. In this case, quantization works as a regularization operation and limits overfitting of the training weights. By reducing the precision of numerical values in network parameters, such as weights and activations, quantization reduces model complexity and prevents the network from memorizing noise or outliers in the training data. This reduction in precision introduces a controlled level of noise or approximation error, which helps smooth out decision boundaries and makes the network less sensitive to small variations in the input data. By promoting a more generalized representation of the data, quantization encourages the network to focus on the most relevant features, thereby reducing overfitting. Furthermore, the computational efficiency gained from quantization, such as faster inference and reduced memory requirements, indirectly contributes to regularization by reducing the risk of overfitting that can arise from longer training times or limited training data. Moreover, this study shows that, for this specific anomaly detection problem, 5-state quantization is optimal as it is the minimum number of quantized states where accuracy and other performance metrics are consistently very close to those of floating-point synapses. For example, quantization to a lower number of states (2-state and 3-state) could be more hardware efficient; however, the performance is not consistent compared to full precision synapses. This motivates our study of a multistate non-volatile synaptic memory that is capable of maintaining at least five different non-volatile resistance states.

### 6.3 Results for Quantized DW-based Autoencoder

In this section, we evaluate and compare the effectiveness of anomaly detection in three different configurations of autoencoders: one with quantized synapses, another with quantized DW-based stochastic synapses, and the third with full precision synapses.

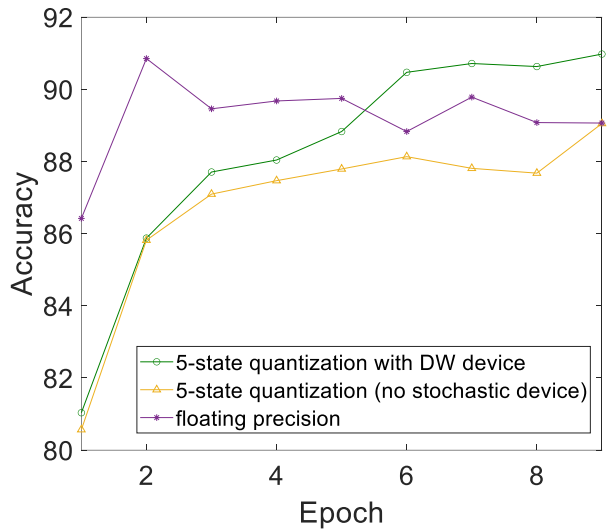

Figure 8: Anomaly detection testing accuracy vs epoch for autoencoder with 5-state quantized DW synapses, 5-state synapses (without device stochasticity), and full precision floating-point weight synapses.

Fig. 8 illustrates anomaly detection testing accuracy for different configurations of the autoencoder. From Fig. 8 it can be inferred that when using only five quantization levels, the quantized autoencoder achieves a competitive accuracy in anomaly detection compared to the autoencoder with full precision floating-point weights. This performance can be attributed to quantization acting as a regularization operation, as explained in the previous section. Additionally, the autoencoder with quantized DW-based stochastic synapses shows higher accuracies than the autoencoder with only quantized synapses. For the autoencoder with DW-based synapses, non-volatile synapses are designed using a specific hardware technology called racetrack MTJ. The synapses can encode multiple non-volatile states and the training process considers the characteristics of the device, such as noise and stochasticity. By introducing randomness during the training process, stochasticity serves as a regularization technique. It adds noise to the model and encourages exploration of different solutions, thus reducing the risk of overfitting to specific patterns in the training data. The results obtained with the DW synapses illustrate a higher anomaly detection accuracy (90.98%) surpassing even the full precision floating-point accuracy (90.85%). The findings presented in Fig. 8 indicate that combining stochasticity with quantization further improves the accuracy of anomaly detection. This combination acts as a better regularization process. Moreover, the fact that stochasticity arises from the inherent properties of the DW devices, rather than being added separately makes it energy efficient as generating random numbers in software can be energy inefficient. Thus, stochasticity inherent to nanoscale devices, which is decremental to Boolean logic, is beneficial to hardware AI applications at no additional energy cost.

**6.4 Total Number of Programmed Weights**

In this section, we conduct a comparison of the total number of programmed weights (weight updates) across different autoencoder synapse schemes.

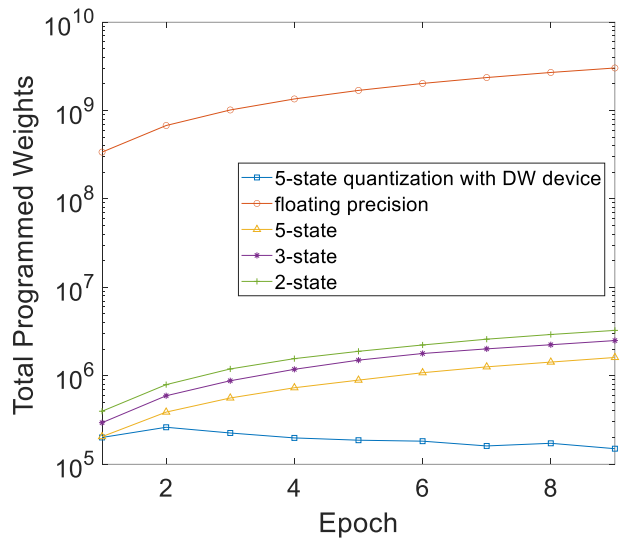

Figure 9: Comparison of the total number of weight updates versus the number of training epochs for different autoencoder types. Significantly fewer weights are updated during the training of the proposed quantized DW device based autoencoder in comparison to the floating precision weight based autoencoder with the same architecture.

Fig. 9 presents the graphical representation of the total weight updates against the number of training epochs for the quantized DW synapse-based autoencoder, as well as the 2, 3, and 5-state autoencoders (excluding stochastic devices), and the floating-point weight-based autoencoder. The data from Fig. 9 reveals a significant distinction: the proposed DW-based approach exhibits a remarkable reduction of at least three orders of magnitude in weight updates when compared to the floating-point approach, implying substantial energy savings for the proposed method. Moreover, the 2, 3, and 5-state quantized autoencoders also demonstrate notably fewer weight updates compared to their floating-point weight-based counterparts. Among these three quantized approaches, the 5-state autoencoder requires the least number of weight updates. Furthermore, it is noteworthy that the number of weight updates for the DW-based autoencoder diminishes as the training epochs progress, in contrast to the patterns observed in the other methods. Consequently, the proposed DW device-based autoencoder demonstrates a greater degree of computational resource efficiency, resulting in reduced energy consumption.

**7. Conclusion**

In conclusion, the state-of-the-art autoencoder based unsupervised anomaly detection methods have shown promising results in detecting network anomalies. However, implementing these methods on edge devices with limited hardware, computational resources, and energy has been a challenge. In this paper, we proposed a solution to this challenge by designing a quantized autoencoder with non-volatile memory-based synapses to detect anomalies on NSL-KDD data effectively on edge devices. Our proposed efficient synaptic weight quantization enables the quantized autoencoder with only 5-state quantization levels to have an anomaly detection accuracy comparable to that of the autoencoder with full precision floating-point weights. Additionally, we designed non-volatile synapses using racetrack MTJ in which the synapses are capable of encoding multiple non-volatile states. The hardware-aware training performed on the 5-state quantized DW based autoencoder yields higher anomaly detection accuracy than the full precision weight autoencoder. Therefore, our proposed solution offers a promising avenue for implementing anomaly detection methods on edge devices with limited hardware resources while being energy efficient. In the future, exploring the compatibility of the proposed quantized autoencoder with diverse datasets and anomaly scenarios could provide valuable insights into its generalizability and robustness. Another potential avenue is the integration of more advanced non-volatile memory technologies into the synapse design, offering the possibility of enhancing the storage and processing capabilities of the autoencoder. Furthermore, the investigation of hybrid models that combine the quantized autoencoders with other anomaly detection techniques could also lead to improved performance outcomes.